\documentclass[aps, a4paper, 11pt, prd, twocolumn, superscriptaddress, showpacs, longbibliography, floatfix, nofootinbib, reprint]{revtex4-2}

\usepackage{bm, amsmath, amssymb, relsize, amsfonts, mathrsfs, multirow, braket, siunitx, color, booktabs, arydshln}
\usepackage{graphicx}
\DeclareSIUnit \parsec {pc}

\usepackage[dvipsnames]{xcolor}
\usepackage[unicode]{hyperref}
\hypersetup{
  colorlinks=true,
  citecolor=cyan,
  linkcolor=red,
  urlcolor=cyan
}

\usepackage{orcidlink}

\newcommand\inp[2]{\langle #1 \,|\, #2 \rangle}

\newcommand{\anstar}[1]{\mathcal{A}_{#1}^{\ast}}

\DeclareMathAlphabet{\mathpzc}{OT1}{pzc}{m}{it}
	
\definecolor{LightCyan}{rgb}{0.88,1,1}
\definecolor{lightgray}{gray}{0.9}

\def \tol       { \texttt{tol} }
\def \match     {\mathscr{M} }
\def \Hz        {\mathrm Hz }

\def \msun      {\rm{M}_\odot}
\def \tlrf      {\mathrm{TaylorF2}}

\def \flow      {f_{\rm low} }

\def \fhigh     {f_{\rm high}}
\def \FF {\mathpzc{FF}}
\def \RF {\mathpzc{RF}}

\def \scipy {\textsc{SciPy}}

\def \dlambda{{\Delta \lambda}}

\def \dmax {\mathcal{D}_{\mathrm{max}}}

\def \IITGn     {Indian Institute of Technology Gandhinagar, Gujarat 382055, India.\vspace*{4pt}}
\def \Nikhef    {\mbox{Nikhef, Science Park 105, 1098 XG Amsterdam, The Netherlands.}\vspace*{2pt}}
\def \Utrecht   {Institute for Gravitational and Subatomic Physics (GRASP), Utrecht University, \mbox{Princetonplein 1, 3584 CC Utrecht, The Netherlands.}\vspace*{3pt}}

\begin{document}

\title{Template bank to search for exotic gravitational wave signals from astrophysical compact binaries}

\author{\textsc{Abhishek~Sharma}\orcidlink{0009-0007-2194-8633}}
\email{sharma.abhishek@iitgn.ac.in }
\affiliation{\IITGn}

\author{\textsc{Soumen~Roy}\orcidlink{0000-0003-2147-5411}}
\email{soumen.roy@nikhef.nl}
\affiliation{\Nikhef}
\affiliation{\Utrecht}

\author{\textsc{Anand~S.~Sengupta}\orcidlink{0000-0002-3212-0475}\vspace*{7pt}} 
\email{asengupta@iitgn.ac.in}
\affiliation{\IITGn}

\begin{abstract}

Modeled searches of gravitational wave signals from compact binary mergers rely on template waveforms determined by the theory of general relativity (GR). Once a signal is detected, one generally performs the model agnostic test of GR, either looking for consistency between the GR waveform and data or introducing phenomenological deviations to detect the departure from GR. The non-trivial presence of beyond-GR physics can alter the waveform and could be missed by the GR template-based searches. A recent study~\cite{Narola:2022aob} targeted the binary black hole merger, assuming the parametrized deviation in lower post-Newtonian terms and demonstrated a mild effect on the search sensitivity. Surprisingly, for the search space of binary neutron star (BNS) systems where component masses range from 1 to  $2.4\:\rm{M}_\odot$ and parametrized deviations span $1\sigma$ width of the deviation parameters measured from the GW170817 event, the GR template bank is highly ineffectual for detecting the non-GR signals. Here, we present a new hybrid method to construct a non-GR template bank for the BNS search space. The hybrid method uses the geometric approach of three-dimensional lattice placement to cover most of the parameter space volume, followed by the random method to cover the boundary regions of parameter space. We find that the non-GR bank size is $\sim$15 times larger than the conventional GR bank and is effectual towards detecting non-GR signals in the target search space.

\end{abstract}
\maketitle

\section{Introduction}
\label{sec:intro}
With the advent of terrestrial observatories such as Advanced LIGO~\cite{LIGOScientific:2014pky}, Advanced Virgo~\cite{VIRGO:2014yos} and KAGRA~\cite{Aso:2013eba}, the gravitational waves (GWs) from compact binary coalescences (CBCs)~\cite{LIGOScientific:2018mvr, LIGOScientific:2020ibl, LIGOScientific:2021djp} allow us to conduct tests of General Relativity (GR)~\cite{LIGOScientific:2016lio, LIGOScientific:2018dkp, LIGOScientific:2019fpa, LIGOScientific:2020tif, LIGOScientific:2021sio}. In addition to GWs, GR has also been tested using solar system observations~\cite{Will:2014kxa}, binary pulsar observations~\cite{Stairs:2003eg, 2014arXiv1402.5594W, Kramer:2021jcw}, and observations of supermassive black holes (BHs) at the center of galaxies~\cite{GRAVITY:2018ofz, Do:2019txf, EventHorizonTelescope:2019ths}. None of them have found any statistically significant deviation yet, which concludes that GR is the most accurate theory of gravity known to date. In particular, compact binary mergers can probe gravity at its most extreme environment characterized by highly dynamical, non-linear and genuinely strong-field regime. Therefore, such systems are exceptional laboratories for unravelling the beyond-GR physics.
\begin{figure*}
\centering
\includegraphics[width=0.98\textwidth]{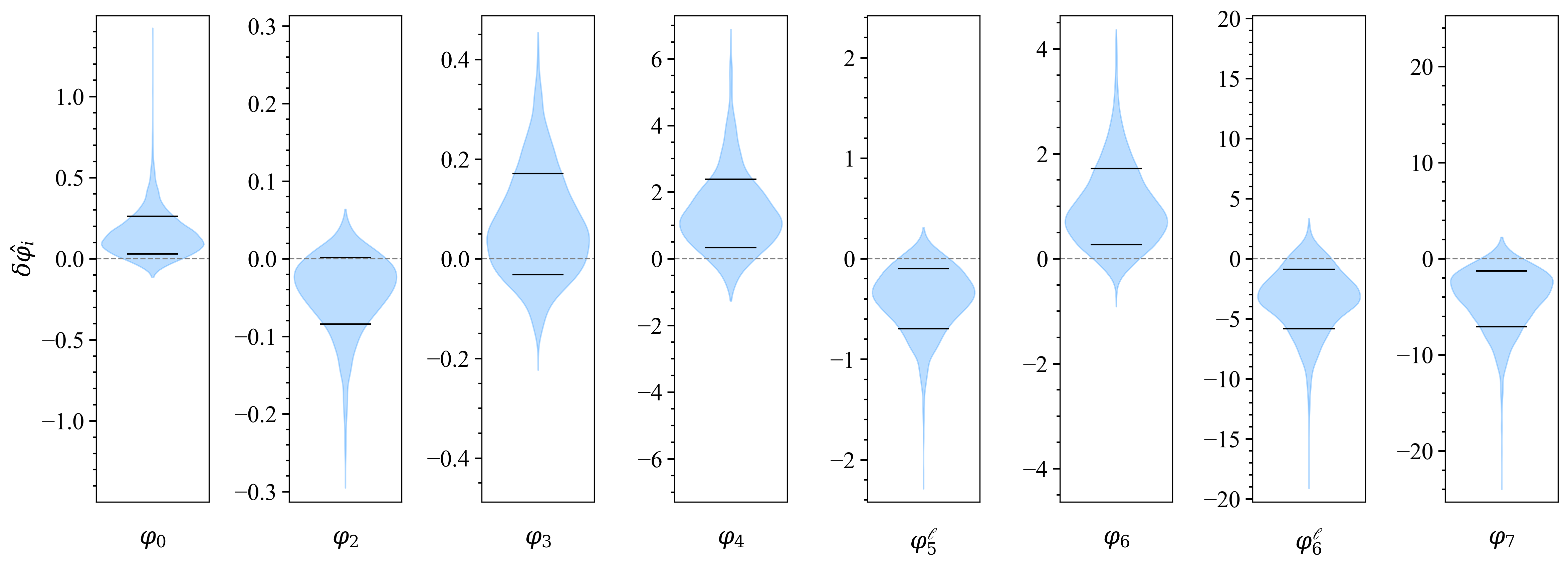}
\caption{Posterior distribution for deviation parameters obtained from the parametrized test with GW170817 event\:\cite{PhysRevLett.123.011102}. Black horizontal ticks mark the $68\%$ credible interval. The grey horizontal dashed line corresponds to the GR prediction.}
\label{fig:deltaPhiRanges}
\end{figure*}

The observations and tests of beyond-GR physics with CBCs can be designed by availing the accurately modeled waveforms for beyond-GR theories. Unfortunately, one lacks a complete understanding of the dynamics of coalescing compact binary in the strong-field regime, in nearly all alternative theories of gravity. Recently, there has been progress toward numerical relativity (NR) simulations of binary black hole mergers in theories beyond GR~\cite{Okounkova:2017yby, Okounkova:2019zjf, Okounkova:2020rqw}. Some of these simulations approximately solve the underlying field equations. In addition to some early developments in NR front, several efforts have been made obtaining the analytical gravitational waveforms in alternative theories \cite{Lang:2014osa, Sennett:2016klh, Bernard:2018ivi, Sennett:2019bpc, Julie:2019sab}. However, a lot more work is still required before these early developments can be incorporated in GW data analysis. Moreover, there might be a more accurate alternative theory, which is unknown to us. Thereby, we generally perform model agnostic analyses: looking for consistency between GR waveform and data~\cite{Cornish:2014kda, Ghonge:2020suv, Roy:2022teu, Ghosh:2016qgn, Maselli:2019mjd}, introducing phenomenological deviations to detect the departure from GR~\cite{Agathos:2013upa, Meidam:2014jpa, Mehta:2022pcn, Saleem:2021nsb}. One example of latter kind of analysis is the {\it parametrized test} of GR, where one measures the deviations in various post-Newtonian (PN) terms of the GR predicted waveform phase.

All these tests are performed {\textit{after}} a GW signal is confidently detected by one of the several search pipelines. The search techniques can be broadly divided into two categories: generic transient searches and template-based searches. The first kind, such as \textsc{cWB}~\cite{Klimenko:2008fu, 2021SoftX..1400678D} and \textsc{oLIB}~\cite{Lynch:2015yin, 2020SoftX..1200620R}, uses minimal assumptions on the GW signature, but are inefficient for long duration signals or when the signal-to-noise ratio (SNR) is low. On the other hand, the template-based searches such as \textsc{GstLAL}~\cite{2017PhRvD..95d2001M, Tsukada:2023edh}, \textsc{MBTA}~\cite{Aubin:2020goo}, \textsc{PyCBC}~\cite{Usman:2015kfa, Nitz:2018rgo, DalCanton:2020vpm} and \textsc{SPIIR}~\cite{Chu:2020pjv}, completely rely on the GR template waveform and are highly efficient for long duration signals. However, a gravitational wave signal carrying a significant amount of non-GR physical effects (in terms of non-zero deviation in the GR-predicted PN phasing coefficient(s)), could be missed by the GR template-based search pipelines.

A recent study~\cite{Narola:2022aob} demonstrated a method for a {\it bottom-up} search for the GW signals that may carry deviations in the PN phasing coefficients from the template waveforms predicted by GR. The study highlighted that the GR bank would fail to detect such a non-GR signal and further showed an improvement in sensitivity for detecting non-GR signals when searched using a non-GR template bank. The study focused on the stellar mass binary black hole (BBH) systems ($m_{1,2} \in [5,\; 50] \;\msun$). The parametrized deviations were considered only in the lower PN terms (upto 2PN) over a range spanning the 90\% credible interval of the posterior distribution reported in the test of GR using events in the GWTC-1~\cite{LIGOScientific:2019fpa}. The fitting factor~\footnote{The fitting factor of a template bank for an arbitrary signal is defined as the maximum value of match over all the templates~\cite{Apostolatos:1995pj}, given in Equation~\eqref{fittingfactor}.} study of GR template bank for non-GR injections revealed that 20\% injections were recovered below the bank's minimal match of 0.97~\cite{Owen:1995tm}. 

\begin{figure}[t]
\centering
\includegraphics[width=0.47\textwidth]{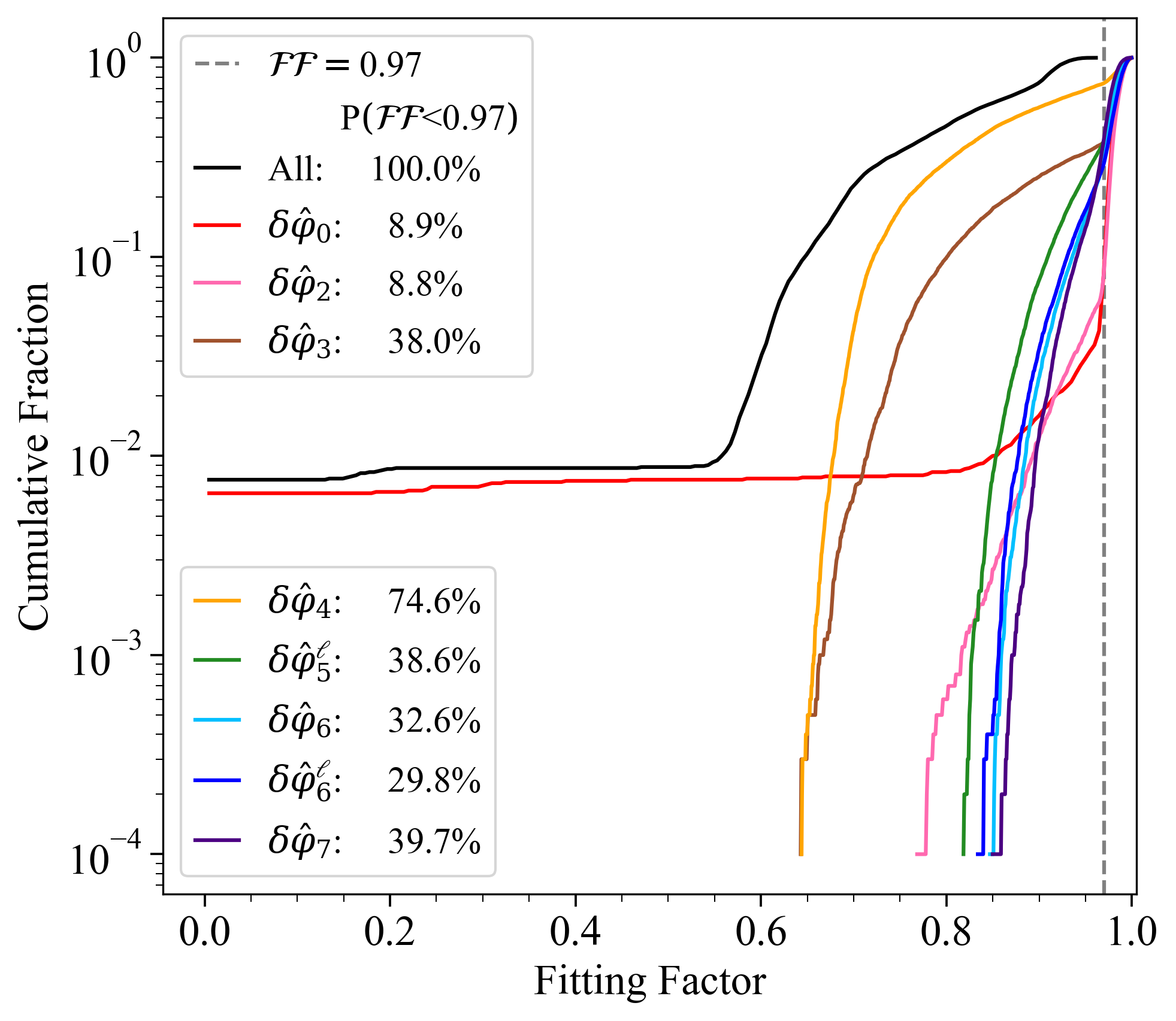}
\caption{Cumulative distribution of fitting factor of GR bank for several sets of non-GR signals, each set comprising $10^4$ signals. The plot is generated using LIGO's O2 sensitivity curve. The GR template bank for BNS search space is generated assuming the mass-spin parameter ranges given in the first two rows of Table~\ref{tab:ParamSpace}, and the minimal match is set to be 0.97 \cite{Owen:1998dk}. Black curve shows the fitting factor for non-GR signals generated assuming deviations at all PN orders. Rest of the curves shows fitting factor for non-GR signals generated assuming deviation in a particular PN phasing coefficient. The deviation parameters are drawn uniformly within the range given in Table \ref{tab:ParamSpace} while generating the non-GR signals. The details of the template bank generation and fitting factor studies are described in Section~\ref{sec:sec3} and Section~\ref{sec:results}, respectively.}
\label{fig:GRvsNonGR}
\end{figure}

In this paper, we target the search space of binary neutron star (BNS) systems with LIGO's sensitivity in the second observing run (O2)~\footnote{Average multi-detector noise PSD: the harmonic mean of the power spectral densities measured from the Hanford and Livingston detectors~\url{https://github.com/gwastro/pycbc-config/tree/65138ade78b234a805e49b694db1c17c20948ecc/O2/psd}} and consider fractional deviations in all of the eight PN terms of the waveform phase up to 3.5PN. Since we are unaware of any astrophysical population in the literature that describes the distribution of parametrized deviations, we consider the  $1\sigma$ (68\% ) credible interval of posterior distribution reported in the test of GR with GW170817\:\cite{LIGOScientific:2018dkp}, as shown in Figure\:\ref{fig:deltaPhiRanges}. These posteriors were obtained by varying only a single deviation parameter at a time, since the multiparameter test leads to an uninformative posterior (see, for example, Figure 7 in \cite{LIGOScientific:2016lio}). However, many alternative theories of gravity possess deviation in more than one post-Newtonian term. For example, assuming stationary-phase approximation (SPA), the leading-order term in the waveform phase
calculated for Brans-Dicke theory appears at -1PN, and the sub-leading terms are known from 0PN up to 1.5PN~\cite{Yunes:2011aa}. 
Therefore, we intend to simultaneously vary all the PN deviation terms to define our non-GR search space. The fitting factor study indicates that the GR template bank for BNS search space is highly ineffectual for detecting the non-GR signals, as shown by the black curve in Figure~\ref{fig:GRvsNonGR}. Non-GR injections are generated by allowing deviations at all PN orders. None of the non-GR injections is found above the desired fitting factor value of 0.97. Suppose a GW signal from a BNS-like system carries a non-GR effect. In that case, it is likely to be missed by the template-based search pipelines and also the unmodeled search pipelines due to their inefficiency for long-duration signals. Figure~\ref{fig:GRvsNonGR} furthermore shows the fitting factor results for the non-GR signals generated assuming deviation in single parameter. For deviations at lower PN order, namely at 0PN and 1PN, approximately 9\% of injections are recovered with a fitting factor below 0.97. Note that for deviation at 0PN, a few injections are recovered with zero fitting factor. For deviation at 1.5PN and higher order PN terms, more than $\sim$30\% injections are recovered with a fitting factor below 0.97. Interestingly, more than 74\% of injections are found below the desired fitting factor for deviation in 2PN.

\begin{table}[ht]
    \centering
   \begin{tabular}{l  l }
    \toprule[1pt]
     \toprule[1pt]
        	\quad  Parameter      			   & \quad  Limits  	 \quad \\
          \midrule[1pt]
         \quad Component masses    &\quad  $m_{1,2} \in [1, \, 2.4] \; \msun$  \quad \\ 
         \quad Component spins       & \quad $ \chi_{1,2} \in [-0.05, \; +0.05 \,]$  \quad \\ 
         \\
         \cdashline{1-2} \\
         \multicolumn{2}{c}{ Deviation parameters } \\
         \\
         \quad 0.0\,PN  & \quad $\delta \hat\varphi_0 \in [+0.029,\; +0.261\,]$ \quad \\
         \quad 1.0\,PN  & \quad $\delta \hat\varphi_2 \in [-0.084,\; +0.001\,]$ \quad \\
         \quad 1.5\,PN  & \quad $\delta \hat\varphi_3 \in [-0.032,\; +0.171\,]$ \quad \\
         \quad 2.0\,PN  & \quad $\delta \hat\varphi_4 \in [+0.329,\; +2.387\,]$ \quad \\
         \quad 2.5\,PN (log-term)  & \quad $\delta \hat\varphi_5^\ell \in [-0.697,\; -0.099\,]$ \quad \\
         \quad 3.0\,PN  & \quad $\delta \hat\varphi_6 \in [+0.266,\; +1.715\,]$ \quad \\
         \quad 3.0\,PN (log-term)  & \quad $\delta \hat\varphi_6^\ell \in [-5.815,\; -0.881\,]$ \quad \\
         \quad 3.5\,PN  & \quad $\delta \hat\varphi_7 \in [-7.067,\; -1.269\,]$ \quad \\
              
\bottomrule[1pt]
\bottomrule[1pt]
    \end{tabular}
    \caption{Parameter ranges used in generating the template bank. The description of parameters below the horizontal dashed line corresponds to the 68\% credible intervals of the deviation in PN coefficients obtained from the parametrized analysis on GW170817 as shown in Figure~\ref{fig:deltaPhiRanges}.}
    \label{tab:ParamSpace}
\end{table}

The fitting factor studies indicate that GW searches using only GR waveforms result in a poor ability to observe BNS systems with non-GR effects. Another approach for detecting GW signals is the unmodeled search, which works without relying on waveform morphology but are inefficient for low-mass binary systems due to their longer in-band duration. As a result, it becomes imperative to develop a beyond-GR search framework, aiming not only to detect new signals but also to explore new aspects of physics.

A crucial input for the matched-filtering search framework is a template bank, which is defined as a collection of theoretical waveforms representing the GWs expected from compact binary mergers. In previous studies, several methods have been proposed to generate a template bank for searches of GW signals from compact binary mergers. The template placement methods are broadly divided into three categories based on their treatment of placement algorithm: geometric, stochastic, and hybrid. Geometric method involves placing templates in the parameter space following a regular, grid-like structure~\cite{Owen:1995tm, Prix:2007ks, Cokelaer:2007mv, Brown:2012qf, Roulet:2019hzy, Hanna:2022zpk, Allen:2021yuy}. This method relies on a parameter space metric that quantifies the mismatch between templates, aiming for uniform coverage. It's efficient for covering a flat parameter space where the metric can be accurately defined. With this method, one can create the most optimal template banks by employing $\anstar{n}$ lattice, which is the best lattice-covering in dimensions $n \leq 5$~\cite{Prix:2007ks}. This method is extensively used to cover the BNS search space with the $\tlrf$ waveform model~\cite{Brown:2012qf}. However, with the current generation of detectors, the $\tlrf$ model is not reliable for detecting the BBH or neutron star-black hole systems systems due to larger merger-ringdown phase contribution to the signal. The parameter space metric of the full inspiral-merger-ringdown waveform families is not flat. Also, the metric formulation for waveforms with precession or higher harmonics is not known to date. This has led to the development of the stochastic (or so-called random placement) method. The stochastic method generates templates randomly over the parameter space and selects those that maximize coverage while minimizing redundancy~\cite{Harry:2009ea, Babak:2008rb, Allen:2022lqr}. As this approach relies on a numerical match function, it is advantageous for exploring complex parameter spaces where the metric is not flat or is unknown. However, its intrinsic random strategy leads to a larger number of templates than needed (so-called overcoverage) and is also computationally intensive. The recent efforts have been focused on developing a hybrid method by combining the space efficiency of the geometric approach with the robustness of the stochastic method\cite{Roy:2017qgg, Roy:2017oul, 2018cosp...42E2899R, Capano:2016dsf, 2017arXiv170501845D, Indik:2017vqq}. The hybrid method is excellent in creating the optimal template banks for the inspiral-merger-ringdown (IMR) waveform families which model the gravitational wave signal from the inspiral, merger and ringdown stages of the dynamics of a compact binary system, and at the same time, less computationally expensive as compared to the stochastic method. And that has led to its widespread use in the LIGO-Virgo-KAGRA collaboration~\cite{LIGOScientific:2018mvr, LIGOScientific:2020ibl, LIGOScientific:2021djp}. This method has been thoroughly developed for non-precessing binaries; extending them to include eccentricity, higher-order modes, or spin precession effects would greatly enhance searches with future observatories.

The GR search space is characterized by four parameters: the component masses and dimensionless spins of the two components of the binary. In contrast, the \mbox{non-GR} search space is characterized by twelve parameters: the four parameters from the GR model plus eight additional deviation parameters in the PN coefficients (up-to 3.5~PN order). A straightforward way to place the templates in twelve-dimensional (henceforth, written as 12-D) non-GR parameter space is to use the stochastic approach. However, creating a stochastic template bank for a higher dimensional parameter space with a larger volume is computationally challenging. In a previous study~\cite{Narola:2022aob}, an 8-D template bank for non-GR BBH parameter space was indeed constructed using the stochastic approach. It was feasible because the targeted search space volume was small and only four PN deviation parameters (upto the 2PN order) were considered.

In this paper, we introduce a new method to create a non-GR template bank for BNS-like systems where we allow for deviations in all the PN coefficients. Following a study by Brown et al.~\cite{Brown:2012qf}, we choose the eight PN phasing coefficients themselves to define the coordinates of an 8-dimensional parameter space, where the parameter space metric becomes flat by construction. Employing principal component analysis, we find that the search space for non-GR signals is effectively three-dimensional embedded within this aforementioned eight-dimensional parameter space. To construct a template bank, we start by placing a geometric grid over the effective 3-D space using $\anstar{3}$ lattice. This geometric grid is further refined by the stochastic placement\footnote{We use the top-down part of the hybrid-geometric random template placement method~\cite{Roy:2017qgg}} to ensure adequate coverage across the entire parameter space. Finally, the points in effective 3-D space are mapped back to the 12-D physical search space using a brute-force method as explained in Sec.~\ref{sec:sec3}. A companion flowchart outlining our template placement algorithm is illustrated in Fig.\ref{fig:flowchart}.

The rest of this paper is organized as follows: in Section~\ref{sec:sec2}, we motivate and describe the search parameter space and the choice of waveform model used to construct the template bank; in Section~\ref{sec:sec3}, we describe in detail the method used to construct the template bank to search for exotic gravitational wave signals; in Section~\ref{sec:results}, we demonstrate the results of validation studies performed to quantify the effectualness of the template banks. Finally, we summarize, conclude and discuss possible future directions in Section~\ref{sec:Conclusion}.

\section{Waveform Model And Search Parameter Space}
\label{sec:sec2}
Neutron stars are formed from the collapse of much heavier stars and in order to conserve angular momentum during collapse, the neutron stars are bound to have large spin at their birth. However, the spinning rate decreases with time due to the magnetic drain of their energy. It is expected to decay away long before entering the band of interest for ground-based gravitational wave observatories~\cite{2004hpa_book}. Consequently, previous studies found that aligned spin template waveforms are sufficient for detecting generic spinning BNS systems with second-generation detectors \cite{Brown:2012qf}. As the post-inspiral and merger signal is emitted at very high frequencies where the second-generation detectors are less sensitive, the signal-to-noise ratio is dominated by the inspiral signal.  Therefore, we only consider the inspiral of the two bodies in this work. The inspiral phase of the waveform can be modeled analytically using the PN framework~\cite{Blanchet:2013haa}. The waveform in the frequency domain can be expressed as,

\begin{equation}
\label{waveform}
  \displaystyle  \tilde{h}(f)=\mathcal{A}(f; D_L, \hat{n}, \vec{\lambda}) \,  \exp\left\{ -i \Psi(f; t_c, \phi_c, \vec{\lambda}) \right\} ,
\end{equation}
where $D_L$ is luminosity distance to the source, $\hat{n}$ describes the sky location ($\alpha, \delta$) and polarization angle ($\psi$) that only affect the overall amplitude and phase of the signal,  $t_c$ is the geocentric coalescence time, $\phi_c$ is the coalescence phase, $\vec{\lambda}$ refers to the set of intrinsic parameters comprising component masses ($m_{1,2}$) and dimensionless spins $(\chi_{1,2})$.
The waveform phase $\Psi(f)$ for $\tlrf$ can be expanded as~\cite{Poisson:1995ef, Buonanno:2009zt},
\begin{equation}
\label{waveformPhase}
\begin{split}
  \displaystyle \Psi&(f)  =  2 \pi f t_c -  \varphi_c - \frac{\pi}{4} \\
  &+  \sum_{k=0}^7 \left[ \varphi_k(\vec{\lambda}; f_0) + \varphi_k^{\ell}(\vec{\lambda}; f_0)\log{x}\right]
  x^{(k-5)/3},
  \end{split}
\end{equation}
where $x=f/f_0$ and $f_0$ is a fiducial frequency. The expansion order $k$ corresponds to ($k/2$)-PN term. Various coefficients at different PN orders are given in Appendix ~\ref{sec:PNCoefficients}.

All the PN phasing coefficients are uniquely determined for given values of the intrinsic parameters. Any deviation from GR would change the binding energy and angular momentum of the binary, thus, transform the equations of binary motion. Alternative theories of gravity will have different functional dependence of PN-phasing coefficients on intrinsic parameters. It has been motivated to devise a parametrized test of GR that works by introducing a fractional deviation parameter ($\delta \hat{\varphi}_i$) for each phase coefficient $\varphi_i$~\cite{Mishra:2010tp, Li:2011cg, Agathos:2013upa, Meidam:2017dgf, Mehta:2022pcn},
\begin{equation}
    \tilde\varphi_i = (1 + \delta \hat{\varphi}_i)\varphi_i^{\rm NS} + \varphi_i^{\rm S},
    \label{parametericDeviations}
\end{equation}
where $\varphi_i^{\rm NS}$ and $\varphi_i^{\rm S}$ are the non-spinning and spin related terms of $\varphi_i$, respectively.

A waveform carrying a nonzero deviation parameter is referred to as a \emph{non-GR} waveform. If a gravitational wave signal carries a significant amount of non-GR physical effects it would be missed by the GR template-based search pipelines as described in ~\cite{Narola:2022aob}, which targeted the BBH search space and allowed deviations in four PN terms from 0.5PN to 2PN order. In this work, we consider the search space of BNS systems and all the deviation parameters up to 3.5PN order. We choose the posterior samples from the parametrized test of GW170817 event studied by LIGO-Virgo-KAGRA (LVK)~\cite{LIGOScientific:2018dkp} and consider $1\sigma$ interval of the marginal posterior distribution of the deviation parameters to define the boundaries of the non-GR space. Parameter ranges for the template bank construction are tabulated in Table~\ref{tab:ParamSpace}.

\section{Template Bank Construction}
\label{sec:sec3}

The long-established method for searching gravitational waves from compact binary mergers relies on the matched filter technique as it is an optimum method to detect a signal if the data contains stationary Gaussian noise. It is accomplished by performing convolution between the data and a set of theoretical filter waveforms to obtain a maximized SNR. In principle, one can maximize the SNR over time, overall amplitude, and overall phase by employing analytical tricks~\cite{Sathyaprakash:1991mt, Dhurandhar:1992mw, Allen:2005fk}. The noise free response of a LIGO-like detector is given by,
\begin{equation}
h(t) = F_+(\alpha, \delta, \psi)\: h_+(t) + F_\times(\alpha, \delta, \psi)\: h_\times(t),
\end{equation}
where $h_+$ and $h_\times$ are two polarizations of gravitational wave, and $F_+$ and $F_\times$ are the antenna response  function for the two polarizations. For dominant $(2, \pm 2)$ mode of gravitational wave signals emitted from nonprecessing binary systems, two polarizations are proportional to each other, $\tilde{h}_+ \propto i \tilde{h}_\times$, where $\tilde{h}_{+,\times}$ denotes the Fourier transform of $h_{+,\times}$. Note that this relation deviates when we include the contribution of higher harmonics or precession to the waveform. When this relation holds, the extrinsic parameters such as the sky location, inclination angle, polarization angle, coalescence phase, and distance to the source can all be expressed as overall amplitude and phase. The intrinsic parameters that can not be absorbed in the analytical maximization procedure, such as component masses and component spins, must be varied to obtain the maximized SNR by generating the filter waveforms repetitively. Our chosen waveform model, $\tlrf$ upholds this relation for both the GR and non-GR waveforms. Therefore, we generate the filter waveforms by varying the intrinsic parameters to obtain the optimal SNR. We evaluate the filter waveforms in the parameter space comprising of component masses and component spins for the GR case and over additional eight deviation parameters for the non-GR case. This discrete set of points in the parameter space constitutes a template bank.

When searching for a signal in data $d$ using a template waveform $h$, the matched filtered SNR is computed by maximizing the inner product between $d$ and $h$ over an overall amplitude, phase ($\varphi_{\rm ref}$), and time ($t_{\rm ref}$),
\begin{equation}
\label{optimalstatistic}
  \mathlarger{\rho}_{\mathrm{\small{MF} }} = \max_{t_{\rm ref}, \varphi_{\rm ref}} \: \inp{ d }{ \hat{h} },
\end{equation}
where $\hat{h}$ is the normalized template waveform such that  $\hat{h} = h/\sqrt{ \inp{h}{h} }$ and the inner product $\inp{\cdot}{\cdot}$ is defined as,
\begin{equation}
 \label{eq:innerProduct}
\displaystyle \inp{a}{b} = 4 \:\Re\int_{\flow}^{\fhigh} \frac{\tilde{a}^\ast(f) \: \tilde{b}(f)}{S_n(f)}\: df,
\end{equation}
where $S_n(f)$ is the one-sided noise power spectral density. Suppose the template waveform does not \emph{exactly} match the signal in the data even after maximizing over the extrinsic parameters. In that case, we will lose a fraction of SNR, which is determined by the mismatch $(1-\match)$ between them. The quantity $\match$ denotes the match between two waveforms:
\begin{equation}
\match(a, b) = \max_{t_{\rm ref}, \varphi_{\rm ref}} \: \inp{\hat{a} }{\hat{b} } 
\label{eq:Match}
\end{equation}

To quantify the effectualness of a template bank ($\mathcal{T}$) for detecting an arbitrary signal $h_a(t)$, we calculate the $\it {fitting \: factor\;}$ ($\mathpzc{FF}$) which is defined as maximal match between arbitrary signal and templates in the bank~\cite{Apostolatos:1995pj},

\begin{equation}
    \mathpzc{FF}(h_a) = \max_{\vec{\lambda} \in \mathcal{T}} \match(h_a, h(\vec\lambda)),
    \label{fittingfactor}
\end{equation}
where $\vec\lambda$ denotes one of the template points.

\subsection{Parameter Space Metric}
\label{sec3:subsec2}
Following~\cite{Owen:1995tm}, the match between two nearby waveforms, whose intrinsic parameters are infinitesimally separated by $\Delta \vec \lambda$, can be Taylor expanded up to the quadratic terms about $\Delta\vec\lambda = 0$ which in turn, can be rearranged to express the \emph{mismatch} $(1-\match)$ in terms of the parameter space metric $g_{ij}$ as,
\begin{equation}
1 - \match \simeq g_{ij} \, \dlambda^i \, \dlambda^j, 
\label{eq:MisMatch}
\end{equation}
where the metric is given by
\begin{equation}
 g_{ij} := -\frac{1}{2}  \left . \frac{\partial^2 \match}{\partial \dlambda^i \: \partial \dlambda^j} \right \vert_{\Delta\vec\lambda = 0}.
\label{eq:metric}
\end{equation}
An alternative approximation for computing the metric is to evaluate the Fisher information matrix of the waveforms over the full parameter space and then project out the dimensions corresponding to the extrinsic parameters~\cite{Owen:1998dk}. The components of the Fisher information matrix are given by,
\begin{equation}
\Gamma_{\alpha \beta} = \left\langle \frac{\partial h(\vec{\theta}) }{\partial \theta_\alpha} \biggm\vert \frac{\partial h(\vec{\theta}) }{\partial \theta_\beta} \right\rangle 
\end{equation}
\begin{figure*}
\centering
\includegraphics[width=0.98\textwidth]{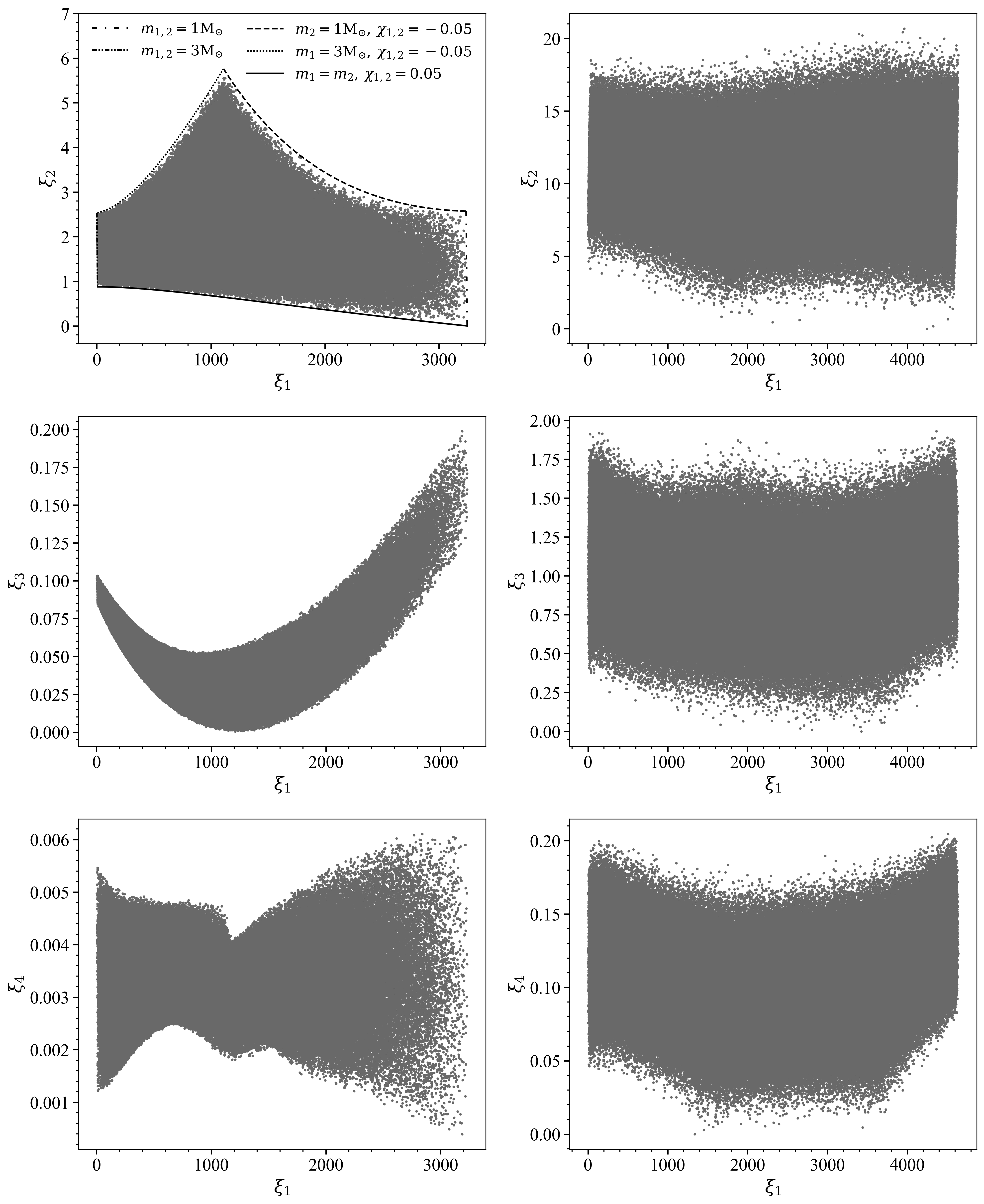}
\caption{The width of the parameter space along $\xi_2, \; \xi_3$ and $\xi_4$ directions, plotted against $\xi_1$ for two distinct cases. Case I (left column): All the deviation parameters ($\delta\hat\varphi_i$) are taken to be zero (GR-case). Case II (right column): All deviation parameters are allowed to deviate from zero and their values lie between the ranges given in Table \ref{tab:ParamSpace} (non-GR case). Component masses and spins are uniformly distributed between the ranges given in Table \ref{tab:ParamSpace} in both cases. Each $\xi_i$ coordinate is scaled such that one unit corresponds to the coverage diameter of $0.97$ minimal match contour, $\dmax$. The dashed curves in the top left plot marks the boundary of GR parameter space in $\xi_1-\xi_2$ plane and corresponding physical parameters are given in the legend. The plot is generated using LIGO's O2 PSD with a lower cut-off frequency of $27\:\Hz$.}
\label{fig:PSextent}
\end{figure*}
where, $\vec{\theta} = \big\{\vec{\lambda},\; \vec{\beta}\big\}$, such that $\vec{\lambda}$ and $\vec{\beta}$ denote the intrinsic and extrinsic parameters, respectively. The parameter space metric is a crucial input for constructing a geometric template bank in order to compute the distance between two points. When placing the templates, a suitable coordinate system is looked for in which the metric components are almost constant. Under this scheme, a coordinate system with minimum intrinsic curvature is considered good. Following~\cite{Brown:2012qf, Ian2014}, we use the PN phasing coefficients comprising six $\varphi_i$ and two logarithmic terms $\varphi_i^\ell$ given in Appendix~\eqref{phaseCoefficients} to define the coordinate system. The metric on this 8-D parameter space does not have intrinsic curvature---the metric components are constant for any point in parameter space. As described in ~\cite{Owen:1995tm, Brown:2012qf}, we similarly use Equation \eqref{eq:metric} to first evaluate the metric in 9-D parameter space including the parameter $t_c$ after maximizing inner product \eqref{eq:innerProduct} over $\varphi_c$, this metric is given by,
\begin{equation}
    \gamma_{\alpha \beta}=\frac{1}{2}\left(\mathcal{J}\left[\psi_\alpha \psi_\beta\right]-\mathcal{J}\left[\psi_\alpha\right] \mathcal{J}\left[\psi_\beta\right]\right),
    \label{9Dmetric}
\end{equation}
where, $\psi_{\alpha}$ denotes the derivative of the $\tlrf$ phase with respect to $\Phi_{\alpha} \equiv \{t_c, \;\Phi_{i}\}$, i.e., $\psi_{\alpha} = \partial \Psi/ \partial \Phi_{\alpha}$, where the index $\alpha$ ranges from $0$ to $8$. And $\Phi_{i}$ denotes the eight PN phasing coefficients, i.e., $\Phi_{i} = \{\varphi_0,\, \varphi_2,\, \varphi_3,\, \varphi_4,\, \varphi_{5l},\, \varphi_6,\, \varphi_{6l},\, \varphi_7\}$. The quantity $\mathcal{J}$ is the moment functional of the noise curve~\cite{Poisson:1995ef, Owen:1995tm}, and is defined as follows for a given function $a(x)$,
\begin{equation}
    \mathcal{J}[a(x)] = \int_{x_L}^{x_H} \frac{a(x)x^{-7/3}}{S_n(xf_0)}\: dx \biggm/ \int_{x_L}^{x_H} \frac{x^{-7/3}}{S_n(xf_0)}\: dx
    \label{momentFunctional}
\end{equation}

\begin{figure*}[t]
\centering
\includegraphics[width=\textwidth]{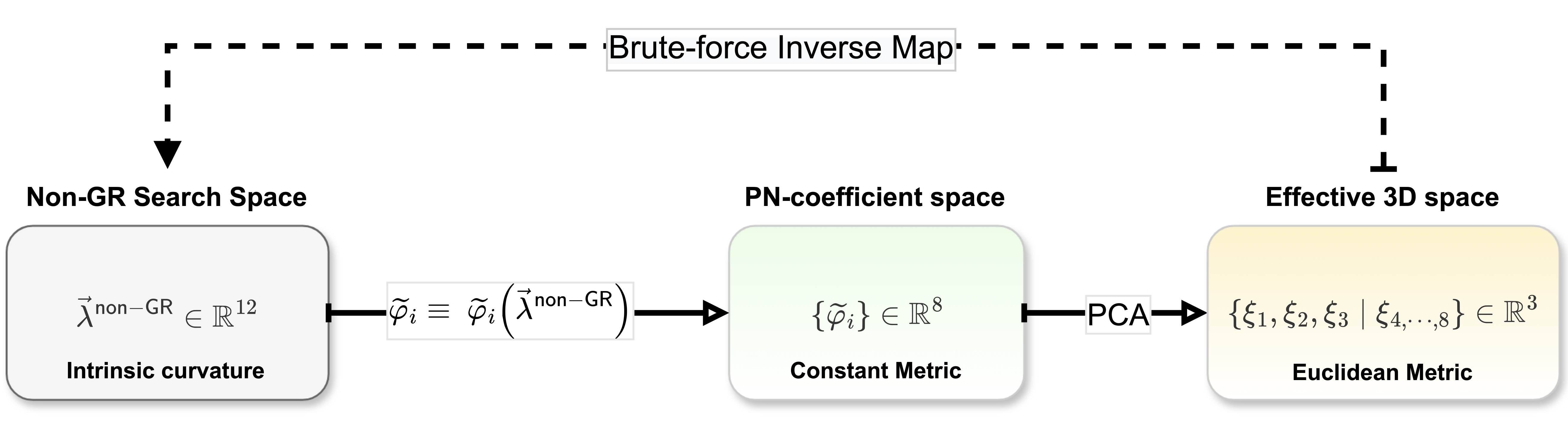}
\caption{Schematic representation of various transformations involved in the construction of the non-GR template bank. The non-GR parameter space is described by the component masses $(m_{1,\,2})$, spins $(\chi_{1,\, 2})$ and eight deviation parameters $(\delta\hat{\varphi}_i)$ and is thus 12-D. Using Eqs.~\eqref{phaseCoefficients} and \eqref{parametericDeviations}, the 12-D vector space is projected to the 8-D space of PN-phasing coefficients, which is flat. Finally, in terms of principal coordinates $\{\xi_i\}\, i=1, \ldots, 8$, the dimensionality is found to be effectively three. The parameter space extent along the rest of the five dimensions $\{ \xi_{4, \ldots, 8} \}$ is sufficiently small to be ignored. Points in this effective 3-D space are mapped back to the physical 12-D non-GR parameter space using a brute force method as explained in Sec. \ref{sec3:subsec4}.}  
\label{fig:flowchart}
\end{figure*}

where $x_L=\flow/f_0$ and $x_H=\fhigh/f_0$ correspond to lowest and highest cutoff frequencies, respectively. The metric $g_{ij}$ on the 8-D subspace, composed of PN-phasing coefficients, is obtained by projecting out the coalescence time $t_c$,
\begin{equation}
    g_{i j}=\gamma_{i j}-\frac{\gamma_{0 i} \gamma_{0 j}}{\gamma_{00}}
\end{equation}
The latin indices $i$ and $j$ range from $1$ to $8$. This projection operation corresponds to the minimization of the distance $\gamma_{\alpha \beta} \Delta \Phi^{\alpha}\Delta \Phi^{\beta}$ with respect to $\Delta t_c$ \cite{Porter_2002}. Since the metric $g_{i j}$ in 8-D parameter space has no dependence on the parameters $(\varphi_k)$ itself, therefore, the parameter space is globally flat in terms of these PN-phasing coefficients.
\subsection{Effective dimensionality of the Parameter Space}
\label{sec3:subsec3}
The eigenvalues of $g_{ij}$ are rapidly decreasing. In particular, the first two eigenvalues are significantly larger than the remaining ones. That indicates the effective dimension of the parameter space must be lower than the dimension of  $g_{ij}$. The extent of the physically relevant region along many directions in the parameter space must be thinner than the maximum mismatch and therefore, we do not need to place templates in those regions of parameter space. 

To identify the effective dimensionality of our parameter space composed of PN-phasing coefficients, we use the principal component analysis-based method proposed in~\cite{Brown:2012qf}.  We first transform to a Cartesian coordinate system by performing rotation and scaling so that the metric becomes the identity matrix. As a result, further rotations will leave the metric unchanged. As $g_{ij}$ is a real-symmetric matrix, its eigenvectors form an orthonormal basis in $\mathbb{R}^8$. The transformation produces a standard basis given by,
\begin{equation}
\label{mu-coordinates}
 \tilde\mu_i = \sum_j \mathcal{R}_{ij} \: \mathcal{S}_{jj}  \: \tilde\varphi_j ,
\end{equation}
where $\mathcal{R}$ is a rotation matrix such that component $\mathcal{R}_{ij}$ is the $j^{\rm th}$ element of the $i^{\rm th}$ eigenvector, and $\mathcal{S}$ is a diagonal scaling matrix, the elements of which are square roots of eigenvalues.

The metric is an identity matrix in this new coordinate system, so we can place the templates using the most optimal $\anstar{n}$ lattice. The width of the parameter space is very thin along many directions, and placing templates in 8-D parameter space would be sub-optimal. Therefore, we perform principal component analysis (PCA) to determine the effective dimension. First, we estimate the covariance matrix in $\tilde\mu_i$ coordinate system by generating a large number of points drawn from uniform distribution within the range of physical parameters as listed in Table~\ref{tab:ParamSpace}, and map them to $\tilde\mu_i$ coordinates using~\eqref{mu-coordinates}. Subsequently, we use eigenvectors of  the covariance matrix to transform from $\tilde\mu_i$ to principal coordinates given by, 
\begin{equation}
    \xi_i = \sum_j C_{i j}\; \tilde\mu_j,
\label{principalCoords}
\end{equation}
where $C_{i j}$ is the $j^{\rm th}$ element of the $i^{\rm th}$ eigenvector. PCA assures that the maximum parameter space extent would lie along the $\xi_1$ direction and the least parameter space extent would lie along the $\xi_8$ direction. 

Figure~\ref{fig:PSextent} illustrates the extent of parameter space in $\xi_i$ coordinates by depicting a large number of random points drawn from the uniform distribution of the physical parameters listed in Table~\ref{tab:ParamSpace}. The figure classifies two cases: the left column shows the extent of GR parameter space where all deviation parameters are set to zero, and the right column refers to the non-GR parameter space that allows deviation parameters to be non-zero. We opt to scale each $\xi_i$ direction such that one unit refers to the coverage diameter of a template, $\dmax=2\sqrt{1-MM}$, where $MM$ stands for minimal match (0.97). This choice is made to visually identify the effective dimensions for placing the templates. For GR parameter space, one can easily notice that the extent along $\xi_3$ and $\xi_4$ directions is smaller than $\dmax$, while the directions $\xi_1$ and $\xi_2$ carry most of the parameter space extent, so a 2-D hexagonal lattice ($\anstar{2}$) would be adequate to construct the template bank. On the other hand, for non-GR parameter space, the extent along the $\xi_3$ is almost twice of $\dmax$, and the extent along $\xi_4$ is five times smaller than $\dmax$. Therefore, we can place $\anstar{3}$ lattice in $\xi_1-\xi_2-\xi_3$ coordinates to cover the non-GR search space.

Although we can easily place the templates in $\xi_i$ coordinates using $\anstar{n}$ lattice, but inverse mapping from $\xi_i$ coordinates to physical parameters has yet to be discovered. On top of it, for the non-GR case we do not know the parameter space boundaries in $\xi_i$ coordinates, which leads to an additional challenge to construct non-GR template bank. Fig. \ref{fig:flowchart} describes a series of coordinate transformations involved in the construction of non-GR template bank.

\subsection{Finding the lattice points in search parameter space}
\label{sec3:subsec4}
\begin{figure*}[t]
\centering
\includegraphics[width=\textwidth]{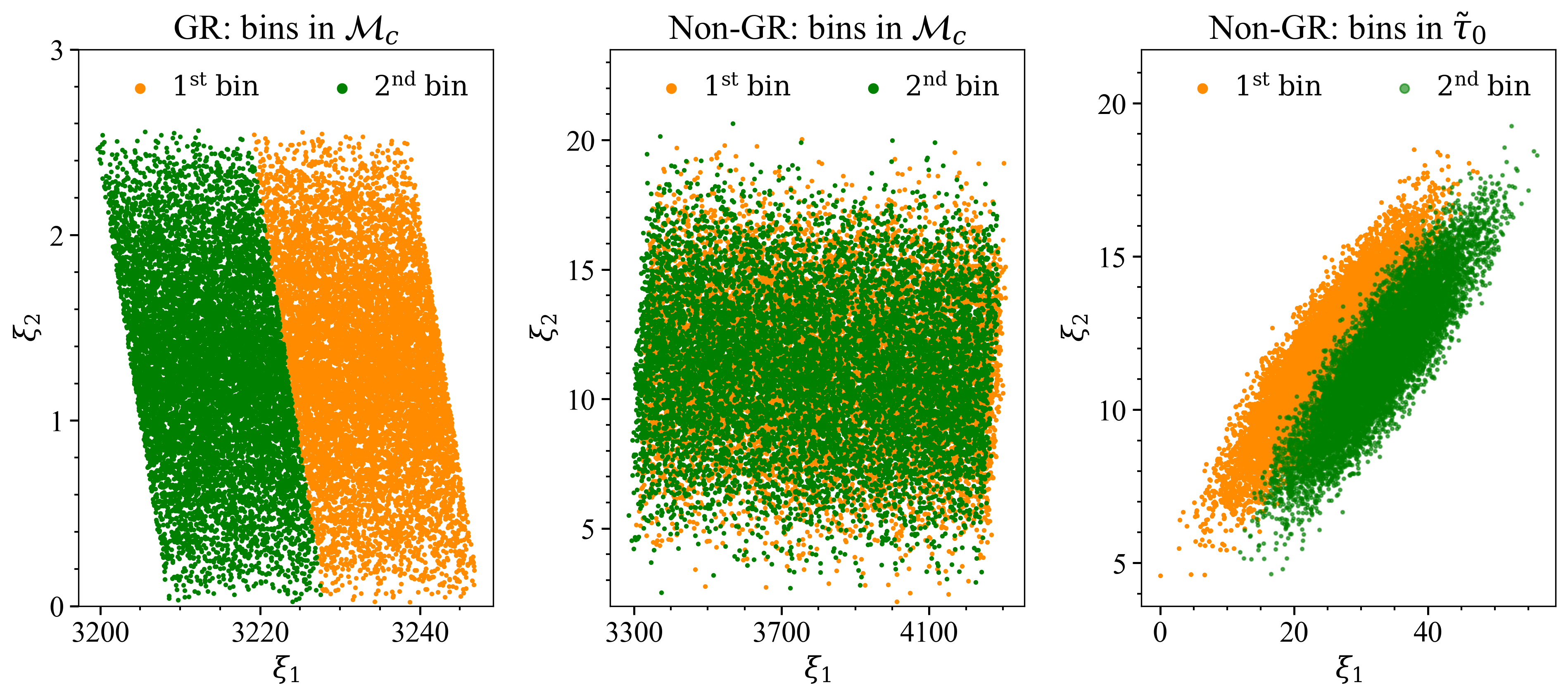}
\caption{An illustration of two consecutive bins over two most leading coordinates $(\xi_1, \xi_2)$, which is used to map the lattice points from $\xi$ coordinate to physical parameter space. The left and middle panel shows the partitioning in chirp mass ($\mathcal{M}_c$) for GR and non-GR space, respectively. The chirp mass binning for non-GR parameter space fails to partition the parameter space due to the strong correlation between $\mathcal{M}_c$ and 0PN deviation term ($\delta\hat{\varphi}_0$). We propose to use the 0PN chirp time ($\tilde{\tau}_0$) as shown in the right panel to partition the non-GR parameter space in $\xi$ coordinates.
}
\label{fig:NNsearch}
\end{figure*}

The GR search space is 4-D, comprising of component masses and spins. The non-GR search space is 12-D due to additional deviation parameters. The template points are generated in $\xi$ coordinates, but there is no inverse mapping to obtain corresponding coordinates in physical search parameter space. Therefore, we follow a brute force method as carried out in previous studies \cite{Brown:2012qf, Ian2014}. For a given lattice point, this method generates random points in the search parameter space and calculates their distance with the lattice point in $\xi$ space. A random point is considered to be a solution when the distance is less than a pre-defined tolerance distance ($\tol$). Throughout this work, we consider $\tol$ to be $10^{-2}$, corresponding to a mismatch of $10^{-4}$ as given in Equation~\eqref{eq:MisMatch}. The volume of a sphere with radius $\tol$ compared to the parameter space volume is $\mathcal{O}(10^6)$ and $\mathcal{O}(10^9)$ times smaller for GR and non-GR cases, respectively. Consequently, finding the solution for all lattice points would be computationally challenging. We alleviate this issue by splitting out the parameter space into non-overlapping subspaces, and using the binary search algorithm KD Tree as implemented in $\scipy$\:\cite{Virtanen:2019joe} to find the nearest random point. The conventional partitioning scheme divides the parameter space over the chirp mass. The left panel of Figure~\ref{fig:NNsearch} shows two consecutive bins over chirp mass ($\mathcal{M}_c = (m_1m_2)^{3/5}/(m_1+m_2)^{1/5}$) for GR parameter space, which are nearly non-overlapping. While for the non-GR case as shown in the middle panel, the two consecutive chirp mass bins are almost entirely overlapping---it's because of the strong degeneracy between chirp mass and 0PN deviation parameter $\delta\hat{\varphi}_0$. Here, we propose to use the 0PN chirp time ($\tilde\tau_0$) that depends on $\delta\hat{\varphi}_0$,

\begin{equation}
    \tilde\tau_0 = \frac{5}{256} \, \mathcal{M}_c^{-5/3}(\pi f_0)^{-8/3} \, (1 + \delta \hat{\varphi}_0) 
\label{eq:NewtonianChirpTime}
\end{equation}
to partition the parameter space in $\xi$ coordinates. The right panel of Figure~\ref{fig:NNsearch} shows that $\tilde\tau_0$ binning reduces the overlap, but it’s less efficient than chirp mass binning for GR space.

We choose those lattice points for which at least one random point is found within $\tol$ to construct a geometric template bank. This bank can cover the bulk region of the parameter space, but the boundaries would not be covered adequately. We use the top-down part of the hybrid geometric-random template placement method\:\cite{Roy:2017qgg, Roy:2017oul, 2018cosp...42E2899R} by seeding the precomputed geometric bank. It starts by generating a large number of random points and then removes the points that are located within the distance of $\dmax/2$ from the existing templates. Later, it picks one point arbitrarily from the remaining random points as a new template and removes those random points that lie at a distance $\leq \dmax/2$ from the chosen point. Continuing this process until all the random points get exhausted generates the hybrid bank.
Figure~\ref{fig:PSextent} shows the random points that are generated assuming uniform distribution over the parameters listed in Table~\ref{tab:ParamSpace}, where fewer random points lie near the boundaries. This is more prominent for non-GR case compared to GR. It happens even if we generate points inside a small bin, as shown in Figure~\ref{fig:NNsearch}. On the other hand, the boundaries of the non-GR parameter space are unknown and one can cover the boundary region by generating considerably larger number of random points, but that would be computationally challenging. A sub-optimal solution is to generate the random points assuming a uniform distribution in non-GR chirp time coordinates ($\tilde\tau_0, \tilde\tau_3$), where $\tilde\tau_3$ corresponds to the non-spinning part of the 1.5PN term,
\begin{equation}
    \tilde\tau_3 = \frac{1}{8 \eta f_0} \, (\pi M f_0)^{-2/3} \, (1 + \delta \hat{\varphi}_3), 
\label{eq:1p5PNChirpTime}
\end{equation}
where $M$ denotes the total mass of the binary ($M=m_1+m_2$).
 
For construction of GR bank, we utilise the known boundaries in $\xi_1-\xi_2$ plane and generate points using stochastic method along the boundaries. We generate 500 bins over $\mathcal{M}_c$ as described above and simultaneously search for nearest random point for all the lattice as well as the boundary points in each bin on different CPU cores. We find that the (hybrid) GR bank contains 21,766 templates, out of which 15,447 corresponds to the $\anstar{2}$ star lattice (geometric GR bank) and rest corresponds to the boundaries. For construction of geometric non-GR bank we generate 1500 bins over $\tilde{\tau}_0$ as described above and search for nearest solution for lattice points in each bin independently on different CPU cores. This bank contains 284,467 templates. Subsequently, in order to provide coverage near boundary region, we initialize the top-down part of hybrid random template placement strategy with 200 million random proposals. 84,067 proposals get accepted as new templates resulting in a hybrid non-GR bank with 368,534 templates.

\begin{figure}[t]
\centering
\includegraphics[width=0.48\textwidth]{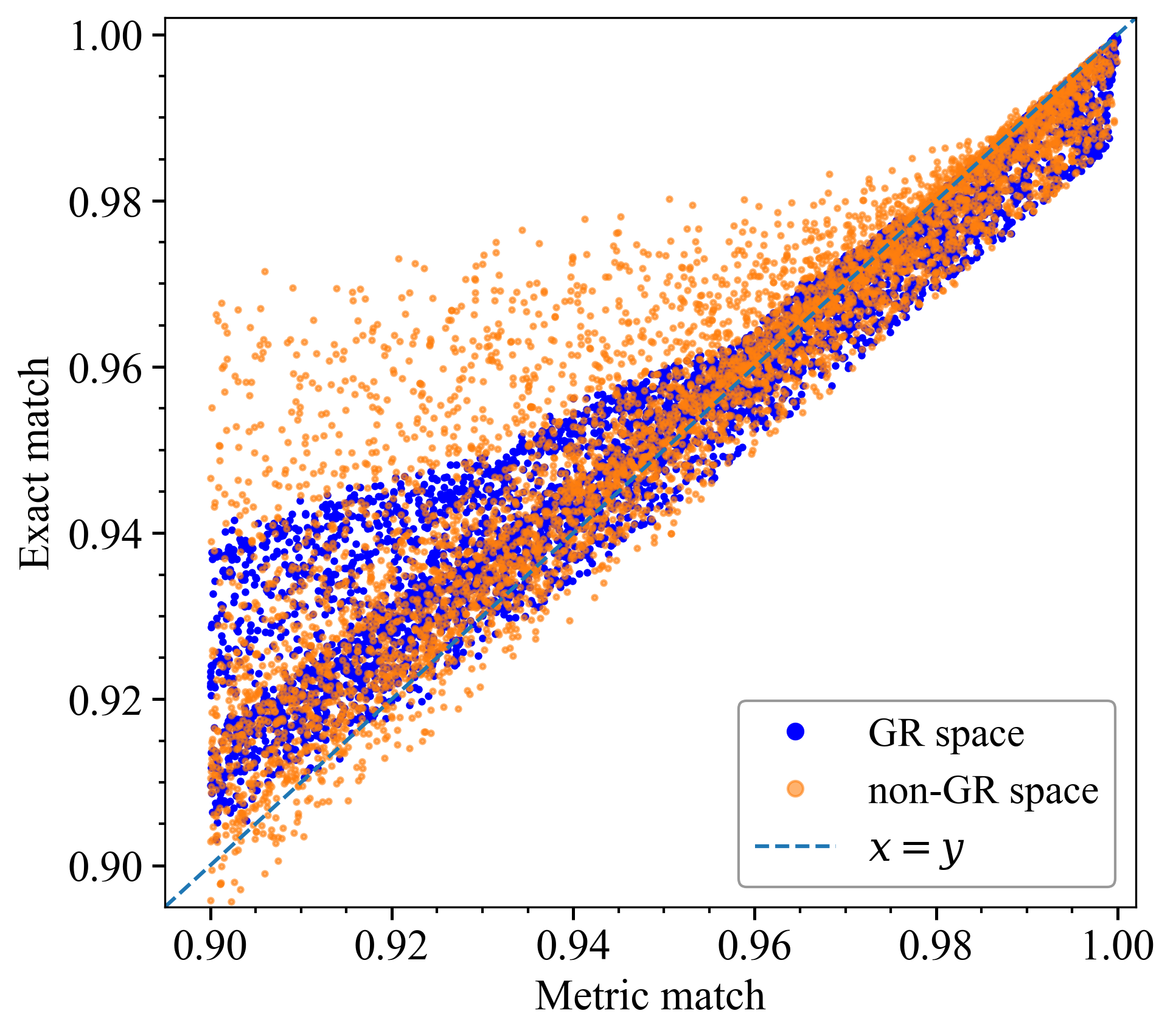}
\caption{A comparison between the match calculated using metric approximation as given by Equation\:\eqref{eq:MisMatch} and the exact match as given by Equation\:\eqref{eq:Match}, shown for two distinct cases: GR (blue dots) and non-GR (orange dots).}
\label{fig:match_comparison}
\end{figure}

\subsection{Failure of metric approximation -- inclusion of exact match}
\label{subsec:exactMatch}
While calculating the over-coverage of the non-GR hybrid bank (see Section~\ref{sec:results}), we inspected the distribution of match between closest pair of templates. More than 22\% of templates are found to have the closest point with a match larger than the considered minimal match, as shown in Figure~\ref{fig:bankRedundancy}. It indicates a non-trivial over-coverage in the bank.

To comprehend the issue of over-coverage, we spray a large number of points in our parameter space and calculate the maximal match of each point with rest of the points using exact match function and the metric (approximate) match as given in Eqs.\:\eqref{eq:Match} and\:\eqref{eq:MisMatch}, respectively. Figure\:\ref{fig:match_comparison} shows the comparison of metric match and the exact match for both GR and non-GR search spaces. For GR search space, we used $\anstar{2}$ lattice to place the templates, where the match between the two closest templates (inter-template match) is expected to be 0.91, but the exact match roughly varies between 0.91 and 0.94. The hybrid method would mostly place the templates near the boundaries, where the inter-template match can reach up to 0.97, and the corresponding exact match varies between 0.96 and 0.975. Consequently, the GR template bank can have over-coverage but not considerable. Similarly, for the non-GR search space, we used $\anstar{3}$ lattice (truncated octahedron) having two kinds of neighbors, one corresponding to square face and other corresponding to hexagonal face (see Appendix B of~\cite{Roy:2017qgg}). The inter-template match values for square and hexagonal faced neighbors are 0.904 and 0.928, respectively. Figure\:\ref{fig:match_comparison} suggests that although corresponding exact match varies across a wide range of values for the two inter-template metric match values, but only a mild fraction of points can have exact match above the minimal match (0.97). However, templates added using hybrid method can have (metric) match values as large as 0.97, where corresponding exact match can even be larger than 0.98 resulting in the observed over-coverage of hybrid non-GR bank. The large disagreement of the metric and exact match while incorporating the deviation parameters indicates the breakdown of metric approximation. 

A recent study~\cite{Harry:2021hls} described a similar breakdown of Fisher matrix approximation for constructing a template bank for BNS systems with tidal deformability. This study demonstrated that including higher-order terms in the Taylor series expansion of the match function can reliably compute the match. However, it is computationally more expensive than the brute-force computation of the exact match and therefore, can not be used in geometric placement. Following~\cite{Roy:2017oul}, we use the exact match function as given in Equation~\eqref{eq:Match} only in the hybrid part of the template placement.

We generate 500 million proposals distributed uniformly in chirp-time coordinates $\{\tilde{\tau}_0, \tilde{\tau}_3\}$, component spins $\{\chi_{1z}, \chi_{2z}\}$ and deviation parameters $\{\delta \hat\varphi_i\}$ within their respective limits as given in Table \ref{tab:ParamSpace} and use minimal match criteria to be 0.965 while constructing hybrid non-GR bank. 52,885 proposals get accepted as new templates. The inclusion of exact match and slightly relaxing the minimal match criteria, resolves the redundancy issue with 8.5\% reduction in bank size. We summarize the size of the geometric and hybrid template banks in Table~\ref{tab:bankResults}.

It takes $\sim$11,500 CPU hours to construct geometric non-GR bank. Generation of the hybrid bank, using geometric bank as the seed, takes $\sim$3500 CPU hours.

\begin{table*}[t]
    \centering
   \begin{tabular}{l c c c c c c c c}
    \toprule[1pt]
     \toprule[1pt]
        Bank       \ \ \ & Type of match \ \  & \ \    Bank size  \ \ &    \multicolumn{2}{c}{\% of $\FF<0.97$ } & \ \ \  & \multicolumn{2}{c}{Lowest $\FF$}  & \ \ Redundancy test \\
          \cmidrule[0.8pt](rr{0.95em}){4-5} \cmidrule[0.8pt](rr{0.95em}){7-8} 
          & computation  & & \ \   GR &  non-GR & &  GR \ \ &  non-GR & \ \  \% of $\RF >0.97$ \\
          \midrule[1pt]
         
          Geometric GR & metric        &  15\,447    &   $1.73$  &  --   & &  $0.948$  & -- & -- \\

          Hybrid GR   & metric         &   $21\,766$  &   $1.26$   &  $100$  & &  $0.962$ & -- & 3.9\\
         
          Geometric non-GR  & metric      &  $284\,467$  &   --  &  $0.52$   & & -- & $0.917$ & 0.0 \\
        
          Hybrid non-GR  & metric &  $368\,534$  &  --  &  $0.17$  & & -- & 0.95 & 22.5\\

          Hybrid non-GR & exact &  $337\,352 $  &   --  &  $0.26$ &   & --  & $0.959$ & 0.0\\
         
    \bottomrule[1pt]
    \bottomrule[1pt]
    \end{tabular}
    \caption{Summary of the GR and Non-GR template banks that are constructed for BNS systems assuming the parameter ranges tabulated in Table \ref{tab:ParamSpace} using $\tlrf$ waveform with O2 PSD, and the lower cut-off frequency is set to be $27\:\Hz$. The 4th and 5th columns report the results from bank validity---how much percentage of the injections are found with a fitting factor ($\FF$) below the desired minimal match value of 0.97. The GR bank is highly ineffectual in recovering non-GR signals as none of the injections is recovered above 0.97.}
    \label{tab:bankResults}
\end{table*}

\begin{figure}[t]
\centering
\includegraphics[width=0.48\textwidth]{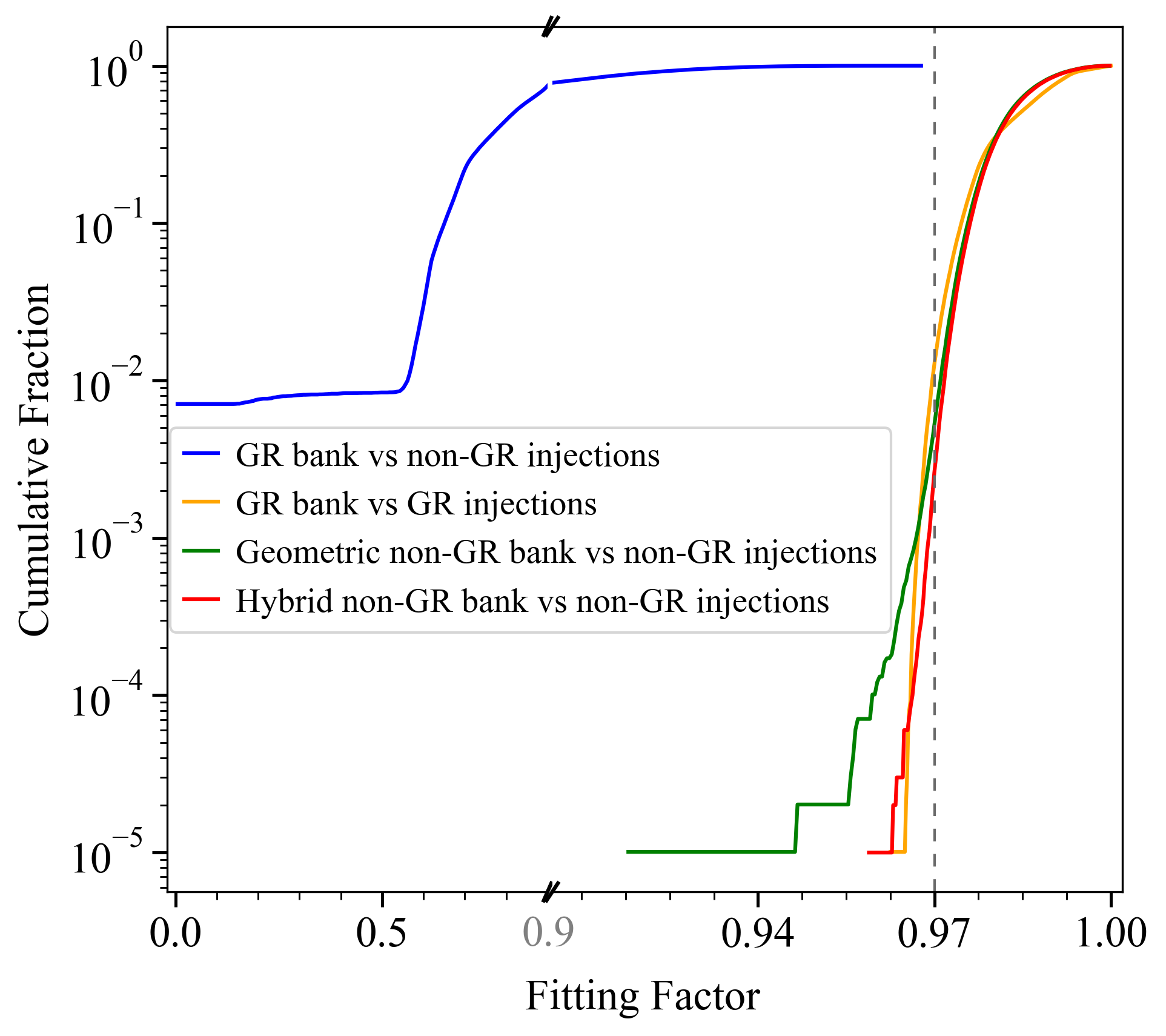}
\caption{Effectualness of the template banks for a set of $10^5\; \tlrf$ injections generated randomly within their target search spaces described in Table~\ref{tab:ParamSpace}. We note that for the geometric non-GR bank, although only $0.52 \%$ of injections are recovered with a fitting factor below the minimal match of the bank $(0.97)$, but it drops to $0.917$ for the worst fitting injection. The complete description of the performance of the banks is listed in Table~\ref{tab:bankResults}. Note that the broken x-axis (splitted at 0.9) is used to accommodate all the Fitting factor distributions together.}
\label{fig:bankEffectualness}
\end{figure}

\begin{figure}[t]
\centering
\includegraphics[width=0.48\textwidth]{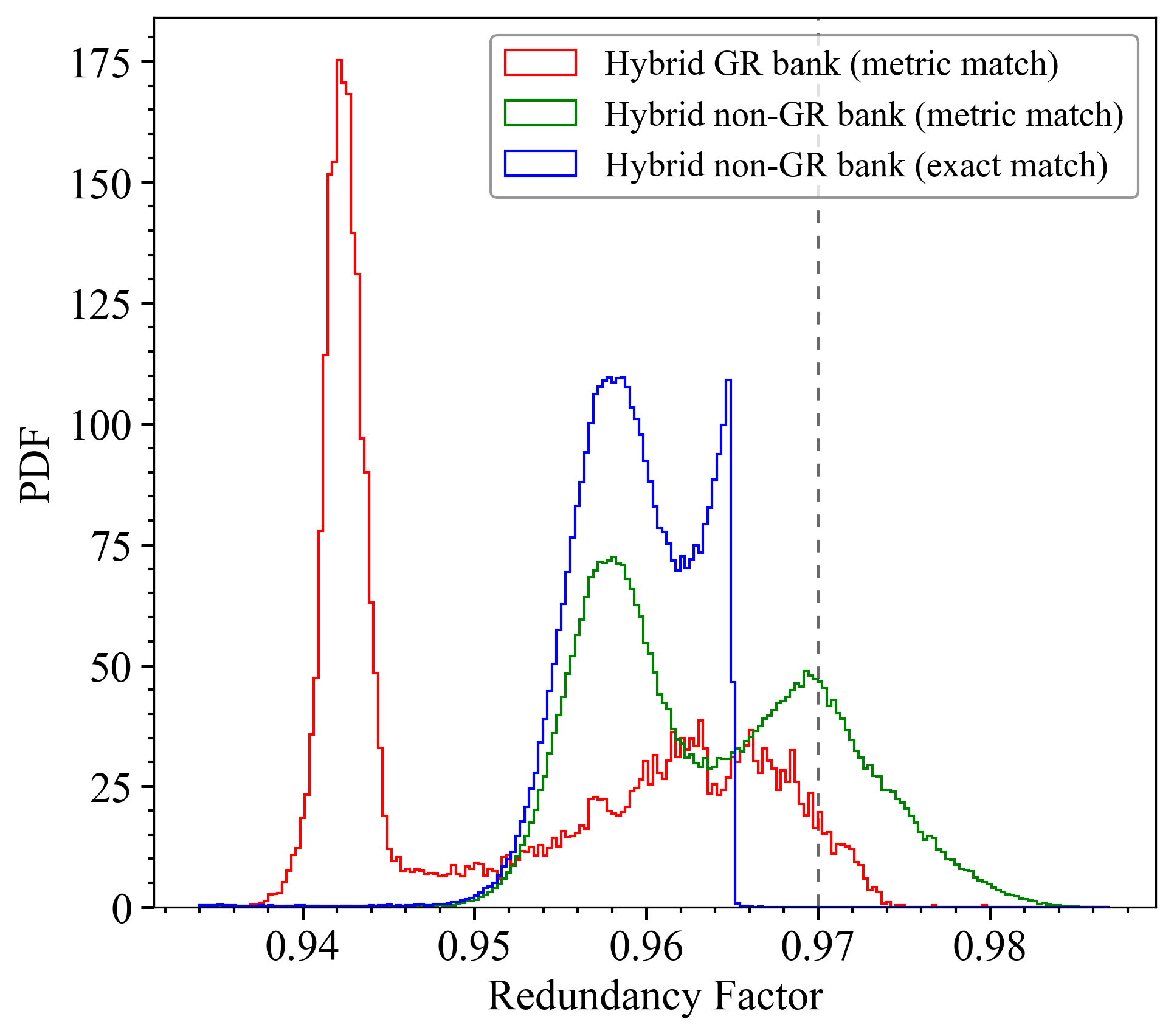}
\caption{Redundancy test of GR as well as non-GR banks. Here, the redundancy factor is calculated for every template against the rest of the templates in the bank. The hybrid non-GR bank with metric match indicates a significant overcoverage, where 22.5\% templates are found above the minimal match of the bank (0.97). The complete description of the redundancy test of the banks is listed in Table~\ref{tab:bankResults}.}
\label{fig:bankRedundancy}
\end{figure}


\section{Template bank validity and redundancy}
\label{sec:results}

To quantify the performance of the template banks, we carry out Monte Carlo simulations to calculate the distribution of the fitting factor for a large set of injections. The fitting factor value for a waveform tells us what fraction of SNR can be recovered, and the distribution can identify the regions of parameter space where the bank coverage is poor. We generate $10^5$ signals using the $\tlrf$ waveform model. The intrinsic parameters are drawn from a uniform distribution within their respective boundaries listed in Table~\ref{tab:ParamSpace}. The fitting factors for non-precessing binary systems are independent of the extrinsic parameters: the sky location, polarization angle and inclination angles. That criteria also holds for non-GR signals. While calculating the fitting factor using Equation~\eqref{fittingfactor}, we generate both the template and injection waveforms with a fixed lower cutoff frequency of $27\:\Hz$. 

Figure~\ref{fig:bankEffectualness} shows the fitting factor distribution of the GR template bank for recovering GR as well as non-GR signals. We note that with GR bank, \emph{all} the non-GR injections are recovered below the fitting factor of 0.97, which implies that the GR template bank is highly ineffectual for detecting the non-GR signals. Figure~\ref{fig:bankEffectualness} also shows the fitting factor distributions of the two non-GR banks for recovering non-GR signals. We note that the new hybrid non-GR template bank has recovered \emph{almost all} the injections above a fitting factor of 0.97. We summarize the fitting factor results in Table~\ref{tab:bankResults}.

We perform the redundancy test of a template bank by calculating the maximal match between a targeted template waveform ($h_i$) and every template waveform ($h_j$) in the bank excluding the template in question~\cite{2023arXiv230410340W}, we call it the redundancy factor ($\RF$),
\begin{equation}
  \displaystyle  \RF(h_i) = \max_{\substack{1\leq j \leq n_{\rm T},  \ j \neq i}} \match(h_i, h_j),
    \label{redundancyfactor}
\end{equation}
where $n_{\rm T}$ is the number of templates in the bank. Ideally, the redundancy factor for any template should not be larger than the minimal match used to construct the bank. Figure~\ref{fig:bankRedundancy} shows the distribution of the redundancy factor for different template banks. For the hybrid non-GR bank constructed using metric match, $22.5 \%$ of templates have a redundancy factor larger than $0.97$, which indicates a significant overcoverage. For comparison, we also construct a similar plot for GR bank and find that for this case, only $3.9 \%$ of the templates have the redundancy factor larger than $0.97$.

\section{Discussion and Conclusion}
\label{sec:Conclusion}

In this work, we have investigated the performance of a GR-template bank for searching the non-GR signals from BNS systems where component masses range from 1 to $2.4\;\msun$, and the phenomenological deviation parameters span $1\sigma$ width of the posterior distribution measured from the GW170817 event. With the LIGO's O2 sensitivity, we have noticed that most of the non-GR signals could be missed by the GR template bank. We have presented a hybrid method for constructing a template bank for searches of beyond GR signals. We found that our non-GR bank size is $\sim$15 times larger than the conventional GR bank. We have shown that our new bank is faithful in detecting the non-GR signals in it's target search space, whereas the GR bank could not recover any non-GR signal above a fitting factor value of $0.97$.

The previous study exploring the searches of non-GR signals~\cite{Narola:2022aob} targeted the search space relevant for BBH systems, and the parametrized deviations were considered in the lower PN terms and constructed an 8-D template bank using the straightforward stochastic method. In this work, we target the 12-D search space of BNS systems, including the deviation in all the PN terms up to 3.5PN order, and present a hybrid method by combining the space efficiency of the geometric method and the robustness of the random method. In this work, we restrict ourselves to use only the fractional deviations i.e., the deviation over the non-zero PN coefficients. In theory one can include deviations on the PN terms for which the coefficient value is zero in GR.

This method can be used wherever the TaylorF2 waveform model is applicable, such as searches of subsolar mass (SSM) compact binary, including eccentricity for SSM/BNS and searches of stellar mass binaries with space-based detectors~\cite{2023arXiv230410340W}. 
While we have constructed the non-GR template bank using the TaylorF2 waveform model in this work, our approach can also be used for IMR waveform models, since our algorithm relies on a ‘hybrid’ method of template placement. Under this approach, one starts by placing an initial geometric grid of templates using the TaylorF2 metric, as generating a geometric bank with the IMR model directly is not feasible. This geometric grid can be further refined by a stochastic placement of additional template points using the ‘exact match’ between new random proposal points and the existing templates in the bank.

The deviation range considered in this work is obtained from LVK analysis of GW170817, which was measured assuming the deviation in a single PN phasing coefficient at a time. Considering deviations in all coefficients simultaneously leads to uninformative posteriors due to correlations among deviation parameters. It might be interesting to consider the uncorrelated non-GR parameters obtained from original deviation parameters through principal component analysis to define the non-GR search space\:\cite{2023GReGr..55...55S, Saleem:2021nsb}. However, the study in\:\cite{2023GReGr..55...55S} considered the deviations only up to 2PN, and we are unaware of any multiparameter test in the literature that evaluated the deviations in all the PN terms.

In future work, we intend to conduct searches of non-GR signals from BNS-like mergers with the LIGO and Virgo's data during the first and second observation runs. A single detection of this type of source could reveal a novel formation channel for compact binaries.

\section{Acknowledgements}
We thank Ian Harry for carefully reading the manuscript and for offering several comments and suggestions to improve the presentation and content of the paper. We are highly grateful for the suggestions received from Alex Nielsen, Tito Dal Canton and Thomas Dent. A.S. thanks IIT Gandhinagar for the research fellowship. S.R. was supported by the research program of the Netherlands Organization for Scientific Research (NWO). We acknowledge computational resources provided by IIT Gandhinagar and also thank high performance computing support staff at IIT Gandhinagar for their help and cooperation. We gratefully acknowledge computational resources provided by the LIGO Laboratory and supported by the NSF Grants No.~PHY-0757058 and No.~PHY-0823459. This research has made use of data, software and/or web tools obtained from the Gravitational Wave Open Science Center, a service of LIGO Laboratory~\cite{GWOSC:catalog}, the LIGO Scientific Collaboration and the Virgo Collaboration. The material of this paper is based upon work supported by NSF's LIGO Laboratory, which is a major facility fully funded by the National Science Foundation (NSF).

$\it{Softwares.}$ To obtain the waveforms and PN coefficients, we use the \textsc{LALSimulation} package of the LIGO Algorithms Library (LAL) software suite~\cite{lalsuite}. The fitting factor studies were performed by modifying $\texttt{pycbc\_banksim}$ code implemented in the \textsc{PyCBC} library~\cite{alex_nitz_2023_7713939}. Our analysis utilize Numpy~\cite{Harris:2020xlr}, Scipy~\cite{Virtanen:2019joe} and Matplotlib~\cite{Hunter:2007ouj}.

\clearpage
\appendix

\onecolumngrid

\section{PN coefficients}
\label{sec:PNCoefficients}

Here, we tabulate the PN expansion coefficients of $\tlrf$ waveform phase as mentioned in Equation\:\eqref{waveformPhase}. The waveform phase contains corrections to Newtonian order up to 3.5-PN order in non-spinning and linear spin-orbit effects \cite{Arun:2008kb, Bohe:2013cla} and up to 3-PN order in quadratic spin effects \cite{Bohe:2015ana}.
\begin{subequations}
{
\allowdisplaybreaks
\begin{align}
\varphi_0 &= \frac{3}{128 \eta} (\pi M f_0)^{-5/3}   
\label{phi0} \\
\varphi_2 &= \frac{3}{128 \eta} \left( \frac{3715}{756} + \frac{55\eta}{9} \right) (\pi M f_0)^{-1}   
\label{phi2} \\
\varphi_3 &= \frac{3}{128 \eta}\left\{-16 \pi + \frac{113\delta\chi_a}{3} + \left(\frac{113}{3} - \frac{76\eta}{3}\right) \chi_s\right\} (\pi M f_0)^{-2/3}
\label{phi3} \\ \nonumber
\varphi_4 &= \frac{3}{128 \eta}\bigg\{\frac{15293365}{508032}+\frac{27145 \eta}{504}+\frac{3085 \eta^2}{72}+\left(-\frac{405}{8}+200 \eta\right) \chi_a^2-\frac{405}{4} \delta \chi_a \chi_s \\ 
          & {\ \ \ }+\left(-\frac{405}{8}+\frac{5 \eta}{2}\right) \chi_s^2 \bigg\}(\pi M f_0)^{-1/3} \label{phi4} \\ 
%
\varphi_5^\ell &= \frac{3}{128 \eta}\left\{\frac{38645 \pi}{756}-\frac{65 \pi \eta}{9}+\delta\left(-\frac{732985}{2268}-\frac{140 \eta}{9}\right) \chi_a+\left(-\frac{732985}{2268}+\frac{24260 \eta}{81}+\frac{340 \eta^2}{9}\right) \chi_s \right\}\log(\pi M f_0)
\label{phi5log} \\ \nonumber
\varphi_6 &= \frac{3}{128 \eta}\bigg\{\frac{11583231236531}{4694215680}-\frac{6848 \gamma_E}{21}-\frac{640 \pi^2}{3}+\left(-\frac{15737765635}{3048192}+\frac{2255 \pi^2}{12}\right) \eta+ \frac{76055 \eta^2}{1728}-\frac{127825 \eta^3}{1296} \\ \nonumber
       & {\ \ \ } -\frac{6848}{63} \log 64+\frac{2270}{3} \pi \delta \chi_a+\left(\frac{2270 \pi}{3}-520 \pi \eta\right) \chi_s + \left(\frac{75515}{144}\delta - \frac{8225}{18}\delta\eta \right)\chi_s\chi_a \\ 
       & {\ \ \ } + \left( -480\eta^2 -\frac{263245}{252}\eta + \frac{75515}{288}\right)\chi_a^2 + \left(\frac{1255}{9}\eta^2 - \frac{232415}{504}\eta + \frac{75515}{288}\right) \chi_s^2\bigg\} (\pi M f_0)^{1/3}
\label{phi6} \\
\varphi_6^\ell &= -\frac{107}{42\eta}\log(\pi M f_0) (\pi M f_0)^{1/3} 
\label{phi6log} \\ \nonumber
\varphi_7 &= \frac{3}{128 \eta}\bigg\{\frac{77096675 \pi}{254016}+\frac{378515 \pi \eta}{1512}-\frac{74045 \pi \eta^2}{756}+\delta\left(-\frac{25150083775}{3048192}+\frac{26804935 \eta}{6048}-\frac{1985 \eta^2}{48}\right) \chi_a \\ 
        & {\ \ \ } +\left(-\frac{25150083775}{3048192}+\frac{10566655595 \eta}{762048}-\frac{1042165 \eta^2}{3024}+\frac{5345 \eta^3}{36}\right) \chi_s \bigg\}(\pi M f_0)^{2/3},
\label{phi7}
\end{align}
}
\end{subequations}
\label{phaseCoefficients}
where ${\gamma_E \approx 0.577216}$ is the Euler constant, ${M\equiv m_1+m_2}$ is the total mass of the binary, ${\eta \equiv m_1m_2/M^2}$ is the symmetric mass ratio, ${\delta \equiv (m_1 - m_2)/M}$ is the asymmetric mass ratio, ${\chi_s \equiv (\chi_1 + \chi_2)/2}$ and ${\chi_a \equiv (\chi_1 - \chi_2)/2}$ are the symmetric and asymmetric combinations of the spins. 

\twocolumngrid

%

\end{document}